\documentclass[aps,amsmath,amssymb,prl,twocolumn,showpacs,superscriptaddress]{revtex4}
\usepackage{bm}
\usepackage{epsfig}

\newcommand{\sgn}{\operatorname{sgn}}
\newcommand{\bnabla}{\mbox{\boldmath$\nabla$}}
\newcommand{\paper}{paper }

\begin{document}

\title{Domain Wall Depinning in Random Media by  AC Fields}

\author{A. Glatz}
\affiliation{Institut f\"ur Theoretische Physik, Universit\"at zu
K\"oln, Z\"ulpicher Str. 77, D-50937 K\"oln, Germany}
\author{T. Nattermann}
\affiliation{Institut f\"ur Theoretische Physik, Universit\"at zu
K\"oln, Z\"ulpicher Str. 77, D-50937 K\"oln,
Germany}\affiliation{LPMMH, Ecole Superieure de Physique et de
Chimie Industrielles, 75231 Paris Cedex 05,
                                        France  }
\author{V. Pokrovsky}
\affiliation{Department of Physics, Texas A\&M University, College
Station, Texas 77843-4242}  \affiliation{Landau Institute for
Theoretical Physics, Chernogolovka, Moscow District, 142432,
Russia}

\date{\today}

\begin{abstract}
The viscous motion of an interface driven by an {\it ac external
field} of frequency $\omega_0$ in a random medium is considered
here for the first time. The velocity exhibits a smeared depinning
transition showing a {\it double hysteresis} which is absent in
the adiabatic case $\omega_0 \rightarrow 0$. Using scaling
arguments and an approximate renormalization group calculation we
explain the main characteristics of the hysteresis loop. In the
low frequency limit these can be expressed in terms of the
depinning threshold and the critical exponents of the adiabatic
case.
\end{abstract}

\pacs{75.60.-d, 74.60.Ge}

\maketitle

The driven viscous motion of an interface in a medium with random
pinning forces is one of the paradigms of condensed matter physics
\cite{Fisher98,Kadar98,Sethna+01}.  This problem arises, e.g., in
the domain wall motion of magnetically or structurally ordered
systems with impurities \cite{review.magnetic.wall} or when an
interface between two immiscible fluids is pushed through a porous
medium \cite{Rubio89}. Closely related problems are the motion of
a vortex line in an impure superconductor \cite{Blatter94}, of a
dislocation line in a solid \cite{Ioffe.Vinokur86} or driven
charge density waves \cite{Thorne96}. For a constant external
driving force this problem has been considered close to the zero
temperature critical depinning threshold
\cite{Natter+92,NarFish93,Ertas+94,Chauve+01} and in the creep
region \cite{Ioffe.Vinokur86,Nattermann86}. More recently, the
results of this approach have been used to study hysteresis
effects in magnets subjected to an external force changing {\it
adiabatically} in time \cite{Lyuksutov99,Nattermann01}. It is the
aim of this \paper to develop a description of pinning phenomena
in an {\it ac-field} in the weak pinning limit. As a main result
we find that the zero temperature depinning transition is smeared
and shows a pronounced {\it velocity hysteresis}. The latter has
to be distinguished from the hysteresis of the magnetization which
persists also in the adiabatic case
\cite{Lyuksutov99,Nattermann01}. Using numerical simulations,
scaling arguments and an approximate renormalization group (RG)
calculation we derive scaling laws for the velocity hysteresis at
zero temperature. Also, we discuss the influence of thermal
fluctuations briefly. Despite the fact, that we use the
terminology of driven interfaces, the results apply
correspondingly also to all the other systems mentioned above.

\begin{figure}[h]
\includegraphics[width=0.9\linewidth]{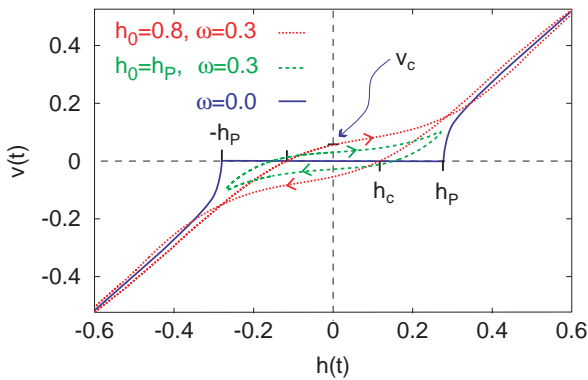}
\caption{Numerical solution of eq. (\ref{eq:motion}) for a 1D
interface in the dc (solid line) and ac case (dotted line) for
$\omega=\omega_0/\omega_P=0.3$ and $h_0 > h_P\approx 0.27$. $x$ is
discretized in $N=1000$ sites and $g$ random in $[-0.5,0.5]$. The
solution is averaged over typically 100 disorder (g-)
configurations. The arrows show the direction of the hysteresis,
for $|h(t)|<h_P$ it is clockwise and anticlockwise for
$|h(t)|>h_P$. The dashed line shows the hysteresis in the case
$h_0\approx h_P$.}\label{fig1}
\end{figure}

{\it Model and zero frequency critical depinning}.--- We focus on
a simple realization of the problem, the motion of a
$D$--dimensional interface profile $z({\bf x},t)$ obeying the
following equation of motion \cite{Feigel83}
\begin{equation}
\frac{1}{\gamma}\frac{\partial z}{\partial t}=\Gamma\bnabla^2z+
h_0\cos{\omega_0t}+g({\bf x},z)\,.\label{eq:motion}
\end{equation}
$\gamma$ and $\Gamma$ denote the mobility and the stiffness
constant of the interface, respectively, and
$h(t)=h_0\sin{\omega_0t}$ is the ac driving force. The random
force $g({\bf x},z)$ is assumed to be Gaussian distributed with
$\big<g\big>=0$ and $\big<g({\bf x},z)g({\bf
x}^{\prime},z^{\prime})\big>= \delta^D({\bf x}-{\bf
x}^{\prime})\Delta_0(z-z^{\prime})$. We further assume
$\Delta_0(z)=\Delta_0(-z)$ to be a monotonically decreasing
function of $z$ for $z>0$ which decays to zero over a finite
distance $l$. Under these conditions the relation between applied
force $h(t)$ and the average velocity $\big<\dot z\big>=v$ shows
in the steady state the {\it inversion symmetry} $h\rightarrow
-h$, $v\rightarrow -v$ (cf. Fig. \ref{fig1}). We therefore
restrict ourselves to the region $h>0$ in the further discussion.

In \cite{Natter+92,NarFish93} eq. (\ref{eq:motion}) was considered
in the adiabatic limit $\omega_0\to 0$. In this case the interface
undergoes a second order depinning transition at $h_0=h_P$ where
the velocity $v$ vanishes as a power law $v\sim (h_0-h_P)^{\beta}$
for $h_0\searrow h_P$, $\beta\le 1$. At $h_0=h_P$ the interface is
self--similar with a roughness exponent $\zeta$, $0\le\zeta <1$,
and the dynamics is superdiffusive with a dynamical exponent $z$,
$1\le z\le 2$. The critical exponents were calculated up to order
$\epsilon=4-D$ in \cite{Natter+92,NarFish93} and recently to order
$\epsilon^2$ in \cite{Chauve+01}, and are related by the scaling
laws $\beta=\nu(z-\zeta)$ and $\nu=1/(2-\zeta)$ \cite{Natter+92},
where $\nu$ denotes the correlation length exponent: $\xi_0\sim
|h_0-h_P|^{-\nu}$. For $h_0\nearrow h_P$ the divergence of $\xi_0$
is related to the increasing size of avalanches.

In the case of an ac-drive the behavior of the system is governed
by the two dimensionless quantities $h_0/h_P$ and
$\omega_0/\omega_P$, where $\omega_P=\gamma h_P/l$. In this \paper
we mainly focus on the most interesting case
$0<\omega_0\ll\omega_P$ and $h_0>h_P$ since this is the region
where universality is expected to hold. As illustrated by the
numerical solution of eq. (\ref{eq:motion}) for $D=1$ at finite
$\omega_0$, the sharp depinning transition is replaced by a
velocity hysteresis, which has clockwise rotation: the velocity
reaches zero at $h(t)=\pm h_c$ for decreasing and increasing
field, respectively (see Fig. \ref{fig1}). For $h_0\gg h_P$ a
second weak hysteresis is found in the region $h>h_c\approx h_P$
which has anticlockwise rotation.

{\it Scaling considerations}.--- First we consider the relevant
length scales of the problem, beginning with $\omega_0=0$.
Comparison of the curvature and the random force term on r.h.s. of
(\ref{eq:motion}) shows, that weak random forces accumulate only
on the Larkin scale $L_P^{\phantom{1}}\approx \big((\Gamma
l)^2/\Delta_0(0)\big)^{1/(4-D)}$ to a value comparable to the
curvature force. On scales $L<L_P$ the curvature force density
$\Gamma l L^{-2}$ is larger than the pinning force density, the
interface is essentially flat and hence there is no pinning. For
$L>L_P$ pinning force densities exceed the curvature forces, the
interface becomes rough and adapts to the spatial distribution of
pinning forces. The largest pinning force density then results
from $L\approx L_P$ from which one estimates the depinning
threshold $h_P^{\phantom{1}}\approx l\Gamma L_P^{-2}$. On scales
$L\gg L_P$ perturbation theory breaks down. The RG calculation
performed in \cite{Natter+92,NarFish93,Ertas+94,Chauve+01}
resulted in a scale dependent mobility and renormalized pinning
forces. A finite frequency $\omega_0$  of the driving force acts
as an infrared cutoff for the propagation of perturbations,
resulting from the local action of pinning centers on the
interface. As follows from (\ref{eq:motion}) (with $\gamma\to
\gamma\left(L/L_P\right)^{2-z}$ for $L>L_P$ \cite{Natter+92})
these perturbations can propagate up to a length scale
$L_{\omega}=L_P(\gamma\Gamma/\omega_0L_P^2)^{1/z}\equiv
L_P(\omega_P/\omega_0)^{1/z}$ \cite{higher.harmonics}: {\it (i)}
If $L_{\omega}<L_P$, i.e., $\omega_0>\omega_P$, $z$ has to be
replaced by 2. During one cycle of the ac drive, perturbations
resulting from local pinning centers affect the interface
configuration only up to scale $L_{\omega}$, such that the
resulting curvature force is always larger than the pinning force
-- there is no pinning anymore and the velocity hysteresis
disappears. {\it (ii)} In the opposite case $L_{\omega}>L_P$,
i.e., $\omega_0<\omega_P$, the pinning forces can compensate the
curvature forces at length scales larger than $L_P$. As a result
of the adaption of the interface to the disorder pinning forces
are renormalized. This renormalization is truncated at
$L_{\omega}$. In the following we will argue, that, contrary to
the adiabatic limit $\omega_0\rightarrow 0$, there is no depinning
transition if $\omega_0>0$. Indeed, a necessary condition for the
existence of a sharp transition in the adiabatic case was the
requirement, that the fluctuations of the depinning threshold in a
correlated volume $\delta h_P\approx h_P
(L_P/\xi_0)^{(D+\zeta)/2}$ are smaller than $(h-h_P)$, i.e.,
$(D+\zeta)\nu\geq 2$ \cite{Natter+92}. For $\omega_0>0$ the
correlated volume has a maximal size $L_{\omega}$ and hence the
fluctuations $\delta h_P$ are given by
\begin{equation}
   \frac{\delta h_P}{h_P}\approx
   \left(\frac{L_P}{L_{\omega}}\right)^{(D+\zeta)/2}=\left(\frac{\omega_0}{\omega_P}\right)^{(D+\zeta)/(2z)}.
   \label{eq:Delta-h_P}
\end{equation}
\begin{figure}[h]
\includegraphics[width=0.7\linewidth]{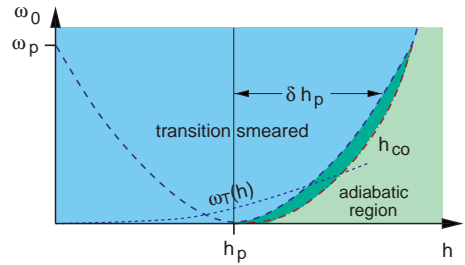}
\caption{Schematic frequency-field diagram for the depinning in an
ac external field (with $h_0>h_P$): For $0<\omega_0\ll \omega_P$
the depinning transition is smeared but traces of the $\omega_0=0$
transition are seen in the frequency dependency of the velocity at
$h=h_P$. This feature disappears for $\omega_0\gg \omega_P$.}
\label{fig2}
\end{figure}
Thus, different parts of the interface see different depinning
thresholds -- the depinning transition is {\it smeared}. $\delta
h_P$ has to be considered as a lower bound for this smearing. A
full understanding of the velocity hysteresis requires the
consideration of the coupling between the different
$L_{\omega}$--segments of the interface, which we will do further
below. When approaching the depinning transition from sufficiently
large fields, $h_0\gg h_P$ ($\omega_0\ll\omega_P$), one first
observes the critical behavior of the adiabatic case as long as
$\xi_0 \ll L_{\omega}$. The equality $\xi_0\approx L_{\omega}$
defines a field $h_{co}$ signaling a cross-over to an {\it inner}
critical region where singularities are truncated by $L_{\omega}$.
Note that $h_{c0}-h_P=h_P(\omega_0/\omega_P)^{1/(\nu z)}\geq\delta
h_P$ (cf. Fig. \ref{fig2}). It is then obvious to make the
following scaling Ansatz for the mean interface velocity
($h_0>h_P$, $v_P=\omega_P l$)
\begin{equation}
   v\left(h(t)\right)\approx v_P
   \left(\frac{\omega_0}{\omega_P}\right)^{\frac{\beta}{\nu z}}
   \phi_{\pm}\left[\left(\frac{h}{h_P}-1\right)\left(\frac{\omega_P}{\omega_0}\right)^{\frac{1}{\nu z}}\right].
   \label{eq:interface_velocity}
\end{equation}
Here the subscript $\pm$ refers to the cases of $\dot h \gtrless
0$, respectively, and $\phi_{\pm}[x\to \infty]\sim x^{\beta}$ (for
$h-h_P\gg h_P$ the classical exponent $\beta=1$ applies). For
$|x|\ll 1$, $\phi_{\pm}$ approaches a constant $c_{\pm}$. Function
$\phi_{-}$ changes sign at a critical value $h_c(\omega_0)\approx
h_P (1-c_{-} (\omega_0/\omega_P)^{1/(\nu z)})$. Direct numerical
solution of eq. (\ref{eq:motion}) in $D=1$ is in good agreement
with this prediction as shown in Fig. \ref{fig3}.
\begin{figure}[h]
\includegraphics[width=0.6\linewidth]{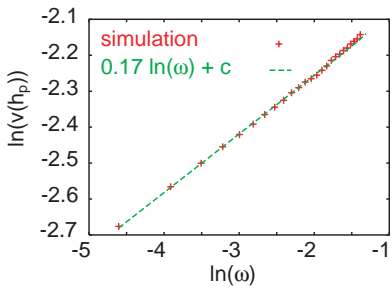}
\caption{Numerical results for $v(h_P)$, using eq.
(\ref{eq:motion}), as a function of $\omega=\omega_0/\omega_P$ in
$D=1$. The dashed line shows the  prediction of eq. (4) with
$\beta/\nu z=0.17$, in good agreement with the value $0.19$ found
in \cite{Chauve+01}.}\label{fig3}
\end{figure}

{\it Renormalized perturbation theory}.--- To consider the
coupling between different $L_{\omega}$-segments we treat model
(\ref{eq:motion}) in perturbation theory. After going over to a
co--moving frame one obtains  in lowest non--trivial order in $g$
the following equation for the velocity $v=\dot z_0(t)$, with
$z_0(t)\equiv\langle z({\bf x}, t)\rangle$ \cite{Feigel83}:
\begin{align}
   \frac{1}{\gamma}v(t) & =
   h(t)+\int_0^{\infty}dt^{\prime}
   \int_{\bf p}\big<\tilde g^{\prime}_{\bf p}
   \big(z_0^{\phantom{1}}(t)\big)
   \tilde g^{\phantom{1}}_{-{\bf p}}\big(z_0^{\phantom{1}}(t-t^{\prime})\big)
   \big>\times\nonumber\\
   & \times\gamma
   e^{-\Gamma\gamma{\bf p}^2t^{\prime}}_{\phantom{1}}
   \equiv h(t)+r_0(t).
   \label{eq:lowest_g}
\end{align}
Here $\int_{\bf p}=\int\frac{d^D{\bf p}}{(2\pi)^D}$ and $\tilde
g_{\bf p}(z)=\int d^Dxe^{i{\bf p}{\bf x}}g({\bf x},z)$. Replacing
the pair correlator of the random forces by
$\Delta_0(z_0(t,t^{\prime}))$ where
$z_0(t,t^{\prime})=\int_{t-t^{\prime}}^t dt^{\prime\prime}\,
v(t^{\prime\prime})$ we get
\begin{equation}
   r_0(t)\sim
   \int_0^{\infty}dt^{\prime}(1+\omega_Pt^{\prime})^{-D/2}
   \Delta_0^{\prime}\big(z(t,t^{\prime})\big).
   \label{eq:r_0}
\end{equation}
which results in corrections to the driving force and to the
mobility which are in general non--local in time. If we assume
that $\Delta_0(z)$ is an analytic function of its argument, it is
easy to show that for $\omega_0\gg \omega_P$, $r_0(t)$ is of the
order $(\omega_P/\omega_0)^2$ and hence small. In this parameter
region the pinning potential merely slows down the motion of the
wall in agreement with the result of our scaling considerations.
In the opposite case, $\omega_0\ll \omega_P$, perturbation theory
breaks down. To treat this frequency region it is instructive to
consider first the case of a {\it dc drive}, $h(t)\equiv h_0$,
where the velocity is constant and hence
$z(t,t^{\prime})=vt^{\prime}$. The $t^{\prime}$--integral
(\ref{eq:r_0}) leads to a correction of the mobility which
diverges  as $(l/v)^{(4-D)/2}$ for $v\to 0$ and $D<4$. This
divergence could be removed by a RG treatment developed in
\cite{Natter+92,NarFish93}. As a result of the elimination of the
Fourier components $z_{{\bf p^{\prime}}}$ with $|{\bf p}|<|{\bf
p^{\prime}}|<L_P^{-1}$ from eq. (\ref{eq:motion}), $\gamma$ and
$\Delta_0(z)$ are replaced there by the renormalized quantities
\begin{eqnarray}
&\gamma(p)\simeq\gamma(pL_P)^{-2+z}\,,&\label{eq:gp}\\
&\Delta_{p}(z)\approx K_D^{-1}(\Gamma l/L_P^{\zeta})^2
p^{4-D-2\zeta}\Delta^{\ast}\left(z(p L_P)^{\zeta}/l)\right).&
\label{eq:replacement}
\end{eqnarray}
$\Delta^{\ast}(x)$ exhibits a cusp-like singularity at $x=0$ which
develops on scales larger than $L_P$. In particular,
$\Delta^{\ast}(x)\approx 1-\sqrt{\epsilon-2\zeta}|x|+
(\epsilon-\zeta)x^2/6+{\cal O}(|x|^3)$ for $|x|\ll 1$
\cite{Natter+92,NarFish93} and
$\Delta^{\ast}e^{-\Delta^{\ast}}=e^{-1-x^2/6}$ for $x\gg 1$
\cite{Fisher86}.  In this way one generates a {\it renormalized }
equation of motion which serves as starting point for a {\it
convergent} perturbative expansion.  The replacements
(\ref{eq:gp}), (\ref{eq:replacement}) are valid for momenta
$\xi^{-1}<p<L_P^{-1}$ where $\xi$ denotes here the correlation
length $\xi_0$ of the zero frequency depinning transition.  In the
spirit of the RG treatment fluctuations on scales larger than
$\xi$ can be neglected since they are uncorrelated. To get the
lowest order corrections in the convergent expansion one has to
replace in eq. (\ref{eq:lowest_g}) the bare quantities by the
renormalized ones. This leads in the limit $v\rightarrow 0$ to
$r_0=h_P\Delta^{\ast\prime}(0^{+})/(2-\zeta)\equiv -\tilde h_P$
which is the RG result for the threshold value $\tilde h_P$.
Replacing in addition $\gamma$ by $\gamma(\xi^{-1})$ on the l.h.s.
of eq. (\ref{eq:lowest_g}), one obtains the correct result for the
critical behavior of the velocity: $v\approx
\gamma(\xi/L_P)^{2-z}(h-\tilde h_P)\approx v_P((h-\tilde
h_P)/\tilde h_P)^{\beta}$.

In the case of an {\it ac drive} the velocity $v(t)$ is periodic
with $2\pi/\omega_0$. In each cycle of $h(t)$ the velocity goes
through a  region of small values, in which perturbation theory
gives a contribution to $\gamma^{-1}$ proportional to
$\big(l/v(t)\big)^{\frac{4-D}{2}}$ as long as the period is large
compared to $l/v(t)$. The cutoff $t_c$ of the
$t^{\prime}$--integration is  given by the  approximate relation
$t_c^{-1}\approx \omega_0+v(t)/l$. These contributions are still
large at $\omega_0, v(t)\rightarrow 0$. This breakdown of
perturbation theory can be overcome by using the RG results
discussed above on intermediate length scales as in the dc case.
Such a procedure is justified for momenta in the range
$L_P^{-1}>|{\bf p}|\gg\xi^{-1}$, where $\xi$ is  the minimum of
$\xi_0$ and $L_{\omega}$. At these scales the interface is still
at criticality.  Since $\xi_0$ depends via $h(t)$ on time and eq.
(\ref{eq:lowest_g}) includes retardation effects, a time dependent
cutoff complicates the problem. Therefore we will restrict our
consideration to the inner critical region where $\xi\approx
L_{\omega}$, i.e., $|h_P-h|<h_P(\omega_0/\omega_P)^{1/\nu z}$. The
renormalized effective equation of motion follows from
(\ref{eq:lowest_g}) with the replacements (\ref{eq:gp}) and
(\ref{eq:replacement}) \cite{largemomenta}:
\begin{eqnarray}
  \frac {v(t)}{\gamma(L_{\omega}^{-1})}& = &
   h(t)+ h_P\omega_P\int_0^{\infty}dt^{\prime}
   \int_{\tilde L_{\omega}^{-1}}^1d\tilde p  \tilde p^{1+z-\zeta}\times\nonumber\\
&& e^{-\omega_P\tilde p^zt^{\prime}} \Delta^{\ast \prime}
(\int_{t-t^{\prime}}^tdt^{\prime \prime} v(t^{\prime\prime})
\tilde p^{\zeta}/l)\,, \label{eq:sce}
\end{eqnarray}
with $\tilde L_{\omega}=(\omega_P/\omega_0)^{1/z}$ and $\tilde
p=pL_P$.  With that form of the equation of motion, we can explain
the hysteresis appearing for $|h|<h_P$ (cf. Fig. \ref{fig1}) more
detailed. First we consider $\dot h<0$: At $h_c$ the sign of the
velocity changes although the driving force is still positive.
This can be understood as follows: until time $t=t_c$, with
$h(t_c)=h_c$, the velocity was positive during half a period,
hence the argument of the $\Delta^{*\prime}$ function is positive.
Therefore the second term of the r.h.s of eq. (\ref{eq:sce}) is
negative and cancels the positive driving force. To solve eq.
(\ref{eq:sce}) analytically we consider a parameter region where
the argument of  $\Delta^{\ast \prime}(x)$ is small compared to
unity, i.e., $\Delta^{\ast\prime}(x)
\approx\Delta^{\ast\prime}(0^{+})\sgn(\int_{t-t^{\prime}}^t
dt^{\prime\prime} v(t^{\prime\prime})) $. One can show a
posteriori that this condition is satisfied if $h_0={\cal O}(
h_P)$. With this approximation the momentum integral in
(\ref{eq:sce}) can be calculated, and we get:
\begin{equation}\label{eq:rgapprox}
\frac{v(t)}{\gamma \tilde L_{\omega}^{2-z}}  \approx
h(t)-\frac{\tilde h_P}{\nu z}\left[S(t,\omega_P)-\tilde
L_{\omega}^{-\frac{1}{\nu}}S(t,\omega_0)\right]
\end{equation}
Here $S(t,\omega)\equiv\int_0^{\infty}
d\tau\,\tau^{-\delta}\tilde\Gamma_{\delta}(\tau)\sgn
z_0(t,\tau/\omega$), $\delta=1/(\nu z)+1$ and
$\tilde\Gamma_{\delta}(\tau)\equiv\Gamma_{\delta}(0)-\Gamma_{\delta}(\tau)$,
where $\Gamma_{\delta}(\tau)=\int_{\tau}^{\infty}dt\, t^{\delta-1}
e^{-t}$. $z_0(t,\tau/\omega)$ changes its sign at
$t=t_0+n\pi/\omega_0$, $n\in\mathbb{Z}$. The dominating part to
$S(t,\omega)$ comes from $\tau<{\cal O}(1)$. To solve this
integral equation for $h\geq 0$ and $\omega_0\ll \omega_P$, we
note, that the sign of $z_0(t,\tau/\omega_P)$ is always positive
for $\tau<1$ and hence $S(t,\omega_P)\approx\nu z$. For $t\lesssim
t_c$ also $z_0(t,\tau/\omega_0)>0$ for the dominating small $\tau$
region of the $\tau$-integration in $S(t,\omega_0)$. This leads to
$\phi_{-}(x)\approx c_{-}+x$, $c_{-}\approx S(t_c,\omega_0)/\nu
z$. By decreasing $t$, $S(t,\omega_0)$ is diminished since regions
with negative $z_0(t,\tau/\omega_0)$ contribute increasingly,
which in turn explains the second weak hysteresis observed in Fig.
\ref{fig1} \cite{Scheidl98}. Next we consider the region $t\gtrsim
t_c$, i.e., $h<h_c$, $v<0$. By increasing $t$, $S(t,\omega_0)$ is
reduced with respect to $S(t_c,\omega_0)$ which leads to a
positive curvature of $v(h)$ for $\dot h<0$. Although that region
is beyond the scope of our RG calculation, since retardation
effects require to consider the avalanche motion in the region
$h<h_c$, it is then tempting to conclude $|v(h=0,\dot h<0)|={\cal
O}(\omega_0/\omega_P)^{\beta/\nu z}$. For large negative values of
$h$, $v(h)$ has to reach again the result of the adiabatic limit.
Together with the inversion symmetry this explains the inner
hysteresis.

{\it Thermal fluctuations}.--- Finally we consider the influence
of thermal fluctuations on the force - velocity relation,
restricting ourselves to the low temperature region $T\ll
T_P=\Gamma l^2 L_P^D$, $T_P$ is a typical pinning energy. {\it
(i)} In the adiabatic limit and for $|h_0-h_P|\ll h_P$, the
velocity obeys the scaling relation
 $v(h,T) = (h-h_P)^{\beta} \psi\left[(h-h_P)^{\theta}/T \right]$
where $\theta$ is a new exponent which depends on the shape of the
potential at the scale $L_P$ \cite{Middleton92}. For $\omega_0>0$
one can extend the scaling relation (\ref{eq:interface_velocity})
to a second scaling field
$\frac{h-h_P}{h_P}\left(\frac{T_P}{T}\right)^{1/\theta}$ and one
finds in particular for $h\approx h_P$
\begin{equation}
v(h_P,T)\approx v_P
\left(\frac{\omega_0}{\omega_P}\right)^{\frac{\beta}{\nu z}}
   \tilde\phi_{\pm}\left[\left(\frac{T}{T_P}\right)^{\frac{1}{\theta}}\left(\frac{\omega_P}{\omega_0}\right)^{\frac{1}{\nu
   z}}\right],
   \label{eq:Tscale}
\end{equation}
with $\tilde\phi_{\pm}[x\rightarrow \infty]\sim x^{\beta}$ and
$\tilde\phi_{\pm}[x\rightarrow 0]\sim\tilde c_{\pm}$. The thermal
smearing of the zero frequency depinning transition is still seen
at finite $\omega_0$ as long as
$\omega_0<\omega_T(h_P)\approx\omega_P\left(\frac{T}{T_P}\right)^{\nu
z/\theta}$. On the other hand for small fields, $h\ll h_P$, the
domain wall shows creep behavior $v(h,T)\approx v_P
e^{-\frac{T_P}{T}\left(\frac{h_P}{h}\right)^{\mu}}$,
$\mu=\frac{2\tilde\zeta+D-2}{2-\tilde\zeta}$, where $\tilde\zeta$
denoted the equilibrium roughness exponent \cite{Ioffe.Vinokur86}.
{\it (ii)} It was shown in \cite{Nattermann01}, that the creep law
is valid also at finite frequencies as long as
$\omega_0\ll\omega_T(h)\approx\omega_P
e^{-\frac{T_P}{T}\left(\frac{h_P}{h}\right)^{\mu}}$, $h\ll h_P$.
For $\omega_0>\omega_T(h)$ and $h_0\ll h_P$ thermal effect are
inessential. Thus, in the region $\omega_0\ll\omega_T(h)$ (Fig.
\ref{fig2}) the force - velocity relation is that of the adiabatic
case at finite temperature.

To conclude, we have shown, that the sharp depinning transition of
an interface driven by a dc field is smeared showing a pronounced
velocity hysteresis, when the external drive is oscillating. The
size of the hysteresis is described by the power laws eqs.
(\ref{eq:interface_velocity}) and (\ref{eq:Tscale}) which are
supported by an approximate renormalization group analysis and a
numerical simulation. The case $h_0<h_P$ will be the subject of
forthcoming studies.

We thank T. Emig, B. Rosenow, S. Scheidl, and in particular S.
Stepanow for fruitful discussions. V. P. acknowledges  support by
NSF under the grant DMR 0072115 and DOE under the grant
DE-FG03-96ER45598, and A.G. and T.N. by Sonderforschungsbereich
608.


\begin{thebibliography}{10}

\bibitem{Fisher98}
D.S. Fisher, Phys. Rep. {\bf 301}, 113 (1998).

\bibitem{Kadar98}
M. Kardar, Phys. Rep. {\bf 301}, 85 (1998).

\bibitem{Sethna+01}
J.P. Sethna, K.A. Dahmen, and C.R. Myers, Nature {\bf 410}, 242
(2001).

\bibitem{review.magnetic.wall}
see, e.g., P.A. Young (ed.) {\it Spin Glasses and Random Fields},
World Scientific, Singapoore 1999.

\bibitem{Rubio89}
M.A. Rubio, C.A. Edwards, A. Dougherty, and J.P. Gollub, Phys.
Rev. Lett. {\bf 63}, 1685 (1989).

\bibitem{Blatter94}
G. Blatter, M.V. Feigel'man, V.B. Geshkenbein, A.I. Larkin, and
V.M. Vinokur, Rev. Mod. Phys. {\bf 66}, 1125 (1994); T. Nattermann
and S. Scheidl, Adv. Phys. {\bf 49}, 607 (2000).

\bibitem{Ioffe.Vinokur86}
L.B. Ioffe and V.M. Vinokur, J. Phys. C {\bf 20}, 6149 (1987).

\bibitem{Thorne96}
R.E. Thorne, Physics Today {\bf 49}, 42 (May 1996).

\bibitem{Natter+92}
T. Nattermann, S. Stepanow, L.-H. Tang, and H. Leschhorn, J. Phys.
II France {\bf 2}, 1483 (1992).

\bibitem{NarFish93}
O. Narayan and D. S. Fisher, Phys. Rev. B {\bf 48}, 7030 (1993).

\bibitem{Ertas+94}
D. Ertas and M. Kardar, Phys. Rev. E {\bf 49}, R2532 (1994).

\bibitem{Chauve+01}
P. Chauve, P. Le Doussal and K. Wiese, Phys. Rev. Lett. {\bf 86},
1785 (2001) and cond-mat/0205108.

\bibitem{Nattermann86}
T. Nattermann, Europhys. Lett. {\bf 4}, 1241 (1986); S. Lemerle et
al., Phys. Rev. Lett. {\bf 80}, 849 (1998).

\bibitem{Lyuksutov99}
I.F. Lyuksyutov, T. Nattermann, and V. Pokrovsky, Phys. Rev. B
{\bf 59}, 4260 (1999).

\bibitem{Nattermann01}
T. Nattermann, V. Pokrovsky, and V.M. Vinokur, Phys. Rev. Lett.
{\bf 87}, 197005 (2001).

\bibitem{Feigel83}
M.V. Feigel'man, Sov. Phys. JETP {\bf 58}, 1076 (1983). We
neglected an inertial term $\rho\frac{\partial^2 z}{\partial t^2}$
which is justified as long as $\gamma\omega_0\rho\ll 1$. A
velocity hysteresis due to inertial effects has been considered by
J.M. Schwarz and D.S. Fisher, Phys. Rev. Lett. {\bf 87}, 096107
(2001).

\bibitem{higher.harmonics}
The consideration of higher harmonics in the interface motion
results in the existence of additional length scales $L_{n\omega}$
which are however of the same scale as $L_\omega$.

\bibitem{Fisher86}
D.S. Fisher, Phys. Rev. Lett. {\bf 56}, 1964 (1985).

\bibitem{largemomenta}
We neglect here contributions from momenta larger then $L_P^{-1}$
which are expected to have a small effect if $\omega_0\ll
\omega_P$.

\bibitem{Scheidl98}
A similar hysteresis loop has seen experimentally in type-II
superconductors, see, e.g., V. Metlushko et al., cond-mat/9804121
(1998).

\bibitem{Middleton92}
A.A. Middleton, Phys. Rev. Lett. {\bf 68}, 670 (1992).


\end{thebibliography}
\end{document}